\documentclass[prd,aps,groupedaddress,showpacs,nofootinbib]{revtex4}
\usepackage{epsf,epsfig,graphicx}

\newcounter{muni}

\begin{document}
\hbadness=10000 \pagenumbering{arabic}
\rightline{DESY-09-100}

\title{Pion transition form factor in $k_T$ factorization}

\author{Hsiang-nan Li$^{1}$}
\author{Satoshi Mishima$^2$}

\affiliation{$^{1}$Institute of Physics, Academia Sinica, Taipei,
Taiwan 115, Republic of China,}

\affiliation{$^{1}$Department of Physics, Tsing-Hua University,
Hsinchu, Taiwan 300, Republic of China,}

\affiliation{$^{1}$Department of Physics, National Cheng-Kung
University, Tainan, Taiwan 701, Republic of China}

\affiliation{$^1$Institute of Applied Physics, National Cheng-Chi
University, Taipei, Taiwan 116, Republic of China}

\affiliation{$^{2}$Theory Group, Deutsches Elektronen-Synchrotron DESY,
22607 Hamburg, Germany}

\begin{abstract}
It has been pointed out that the recent BaBar data on the
$\pi\gamma^*\to\gamma$ transition form factor $F_{\pi\gamma}(Q^2)$
at low (high) momentum transfer squared $Q^2$ indicate an asymptotic
(flat) pion distribution amplitude. These seemingly contradictory
observations can be reconciled in the $k_T$ factorization theorem:
the increase of the measured $Q^2F_{\pi\gamma}(Q^2)$ for $Q^2> 10$
GeV$^2$ is explained by convoluting a $k_T$ dependent hard kernel
with a flat pion distribution amplitude, $k_T$ being a parton
transverse momentum. The low $Q^2$ data are
accommodated by including the resummation of $\alpha_s\ln^2x$, $x$
being a parton momentum fraction, which provides a stronger
suppression at the endpoints of $x$. The next-to-leading-order
correction to the pion transition form factor is found to be less
than 20\% in the considered range of $Q^2$.
\end{abstract}


\maketitle

\begin{figure}[t]
\begin{center}
\includegraphics[height=8.5cm]{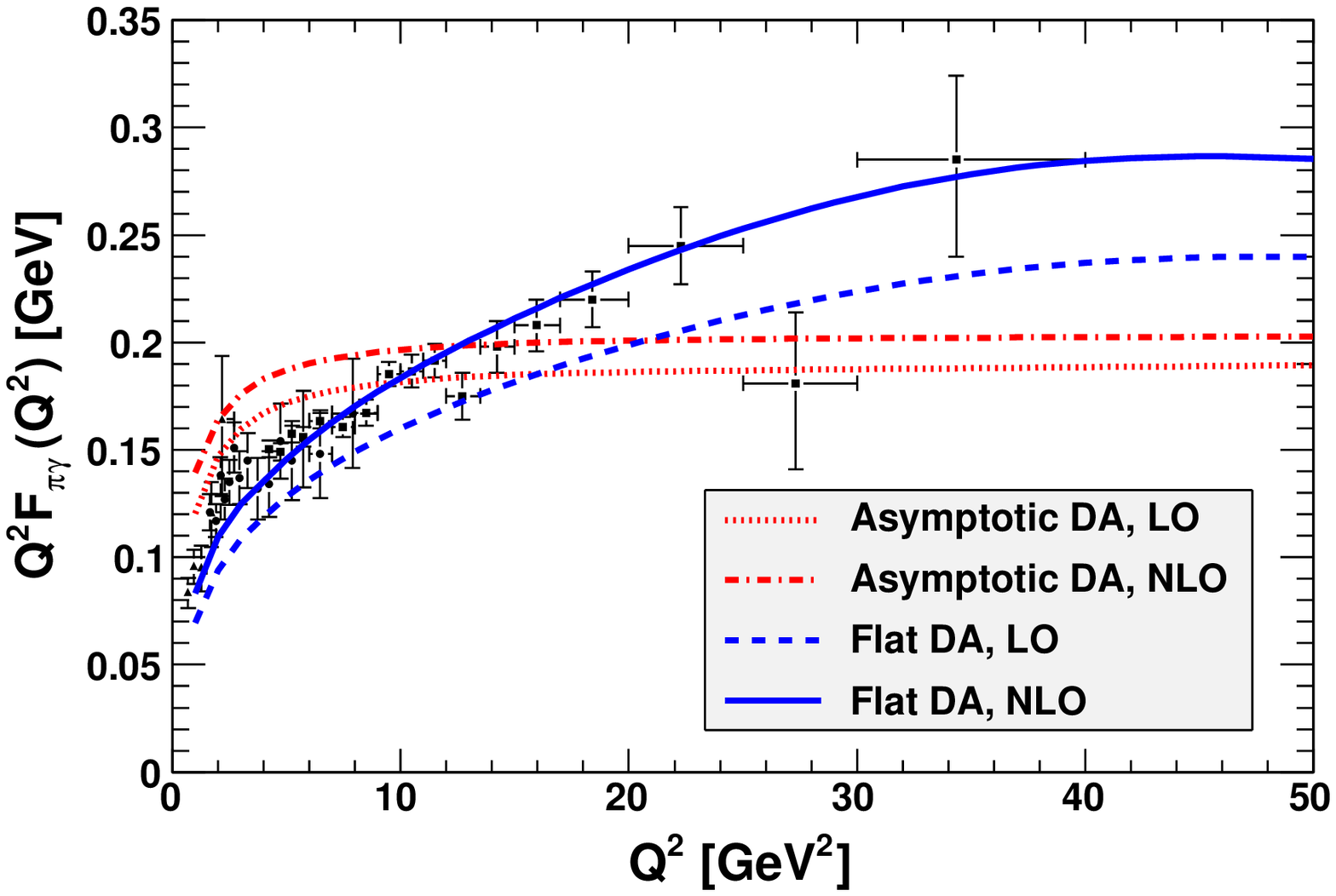}
\vspace{-6mm} \caption{$Q^2$ dependence of $Q^2F_{\pi\gamma}(Q^2)$
up to LO and NLO from the $k_T$ factorization for the asymptotic
(red, dotted and dot-dashed lines, respectively) and flat (blue,
dashed and solid lines, respectively) pion DAs. The points with
errors are the data from CELLO \cite{CELLO}, CLEO \cite{CLEO}, and
BaBar \cite{BABAR}.}\label{fig1}
\includegraphics[height=8.5cm]{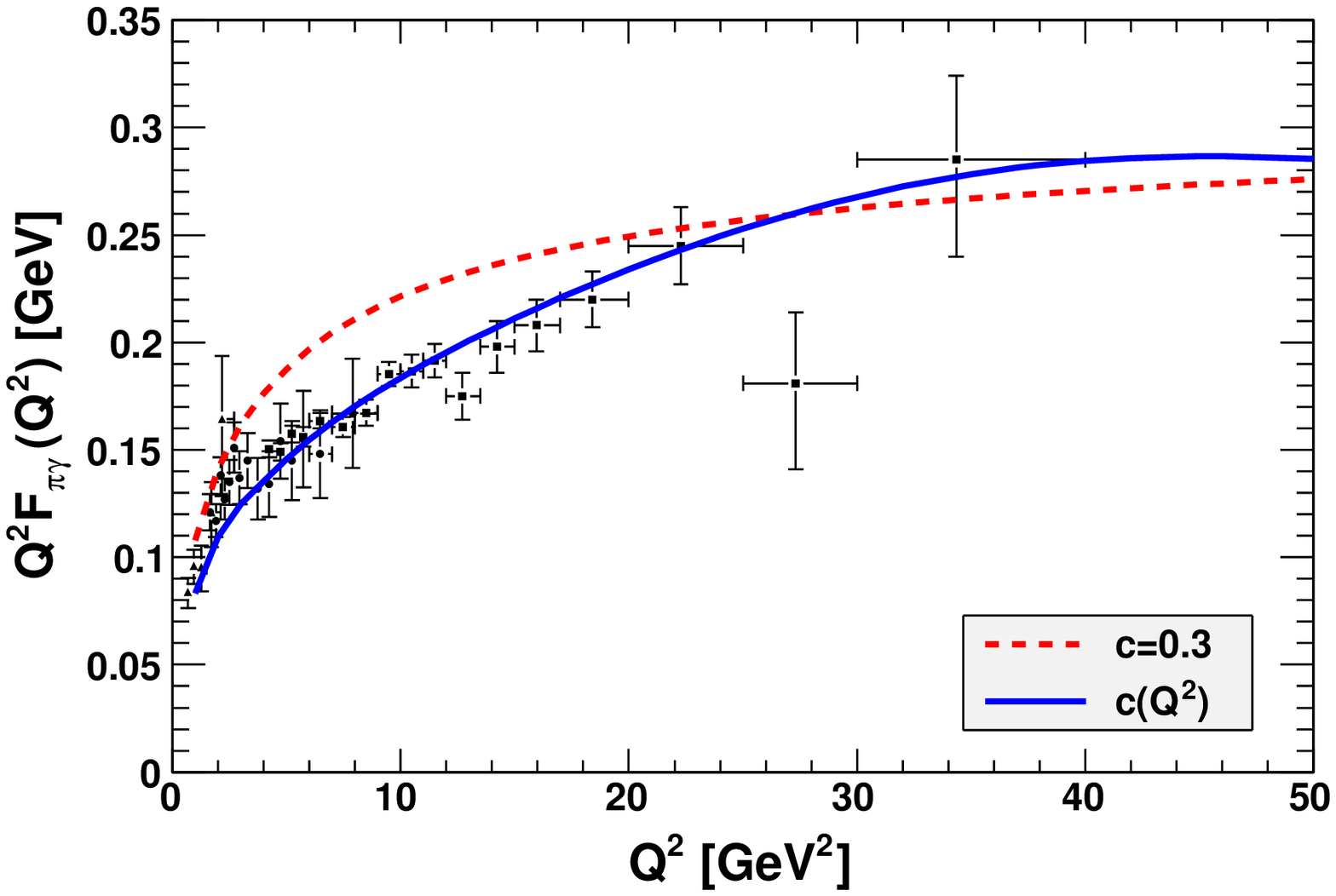}
\vspace{-6mm}
\caption{
$Q^2$ dependence of $Q^2F_{\pi\gamma}(Q^2)$ up to NLO with the power
$c=0.3$ (red, dashed line) and $c(Q^2)$ (blue, solid line) in
Eq.~(\ref{power}).}\label{fig2}
\end{center}
\end{figure}

The asymptotic and soft behaviors of the pion transition form factor
$F_{\pi\gamma}(Q^2)$ involved in the process $\pi\gamma^*\to\gamma$
have been derived \cite{Brodsky:1981rp}:
\begin{eqnarray}
\lim_{Q^2\to\infty} Q^2F_{\pi\gamma}(Q^2)&=&
\sqrt{2}f_\pi\,=\,0.185\;,
\label{as}\\
\lim_{Q^2\to 0} F_{\pi\gamma}(Q^2)&=&
\frac{\sqrt{2}}{4\pi^2f_\pi}\;,
\end{eqnarray}
$Q^2$ being the momentum transfer squared carried by the virtual
photon, and $f_\pi=0.131$ GeV the
pion decay constant. The former is predicted by perturbative QCD
(PQCD) in the collinear factorization theorem
\cite{LB,ER,CZS,CZ,Li:2000hh,Nagashima:2002ia}, while the latter is
determined from the axial anomaly in the chiral limit. However, the
recent BaBar data on $F_{\pi\gamma}(Q^2)$ exhibits an intriguing
dependence on $Q^2$ \cite{BABAR}: $Q^2F_{\pi\gamma}(Q^2)$ exceeds
Eq.~(\ref{as}) for $Q^2>10$ GeV$^2$, and continues to grow up to
$Q^2\approx 40$ GeV$^2$ as shown in Fig.~\ref{fig1}. It has been
commented \cite{Mikhailov:2009kf} that this behavior cannot be
explained by perturbative effects, like higher-order or higher-twist
contributions.

It is known that one can extract nonperturbative information on the
shape of the leading-twist pion distribution amplitude (DA) from the
measurement of $F_{\pi\gamma}(Q^2)$ \cite{LB}. Inspired by the BaBar
data, a model for the pion DA, differing from those investigated
in \cite{Mikhailov:2009kf}, has been proposed
\cite{Polyakov:2009je},
\begin{eqnarray}
\phi_\pi(x,\mu_0)=N+(1-N)6x(1-x),\label{pol}
\end{eqnarray}
with $N$ being a free constant. This model remains finite at the
endpoints of the momentum fraction $x\to 0, 1$ for the normalization
point $\mu_0=0.6\sim 0.8$ GeV. It was argued \cite{Polyakov:2009je}
that the pion DAs from the instanton theory of QCD vacuum
\cite{inst,PP98}, from the the Nambu-Jona-Lasinio model
\cite{RuizArriola:2002bp}, from the chiral quark model
\cite{Bzdak:2003qe}, and from the large-$N_c$ Regge model
\cite{RA:2006ii} are expected to be rather flat. At the same time,
the hard kernel, proportional to the internal quark propagator, was
modified by introducing an infrared regulator $m^2$
\cite{Polyakov:2009je}. Note that an endpoint singularity would be
developed for the model in Eq.~(\ref{pol}) without the infrared
regulator. Then the factorization formula
\begin{eqnarray}
Q^2F_{\pi\gamma}(Q^2)=\frac{\sqrt{2}f_\pi}{3}\int_0^1
dx\,\frac{\phi_\pi(x,Q)}{x+m^2/Q^2}\;,\label{Pol}
\end{eqnarray}
leading to a logarithmic increase $\ln Q^2$,
explains the growth of $Q^2F_{\pi\gamma}(Q^2)$ observed by BaBar
with the tuned parameters $N=1.3\pm 0.2$ and $m=0.65\pm 0.05$ GeV.
The regulator $m^2$ was interpreted as the inverse instanton size,
which sets the scale for nonperturbative effects in a quark
propagator \cite{Polyakov:2009je}.

A similar conclusion on the shape of the leading-twist pion DA has
been drawn from the BaBar data in \cite{Radyushkin:2009zg}, but with
the difference in the formalism that the parton transverse
momentum $k_T$ was taken into account. The hard kernel is the
same as in the collinear factorization, and the $\ln Q^2$
dependence comes from the integration of the pion wave function over
the intrinsic parton $k_T$ up to $xQ$ \cite{Radyushkin:2009zg}:
\begin{eqnarray}
Q^2F_{\pi\gamma}(Q^2)=\frac{\sqrt{2}f_\pi}{6\pi}\int_0^1
\frac{dx}{x}\int_0^{xQ}k_Tdk_T\,\psi_\pi(x,k_T)\;,
\end{eqnarray}
with the model
\begin{eqnarray}
\psi_\pi(x,k_T)=\frac{2\pi\phi_\pi(x)}{x(1-x)\sigma}
\exp\left[-\frac{k_T^2}{2\sigma x(1-x)}\right].\label{flat}
\end{eqnarray}
The cutoff of $k_T$ at $xQ$ guarantees that the integration over $x$
down to $0$ is finite even for a flat distribution $\phi_\pi(x)$,
with which the BaBar data are accommodated by choosing the parameter
$\sigma=0.53$ GeV$^2$. A concern was raised for a broad pion DA in
the conventional collinear factorization \cite{Radyushkin:2009zg}:
the next-to-leading-order (NLO) correction to the hard kernel
\cite{NLO} is huge for the renormalization scale $\mu\sim Q$, or
$\mu$ has to be much smaller than the scale $\Lambda_{\rm QCD}$ in
order to diminish the NLO correction. For the calculation of the
next-to-next-to-leading-order correction to the pion transition form
factor in the collinear factorization, refer to \cite{Melic}.

The BaBar data are consistent with those of CELLO \cite{CELLO} and
CLEO \cite{CLEO} below $Q^2\approx 9$ GeV$^2$, from which an
endpoint suppressed shape of the pion DA has been extracted
\cite{BMS,Guo:2008zzg}. Taking into account
high-power corrections, the asymptotic pion DA can also explain
the CLEO data \cite{Agaev:2004dc}. The QCD sum rule analyses
of the pion DA and
of the pion transition form factor led to an endpoint
suppressed model \cite{Bakulev:2001pa,Radyushkin:1996tb,
Khodjamirian:1997tk,Schmedding:1999ap,
BMS,Agaev:2005rc,Mikhailov:2009kf}. It has been noticed
\cite{JKR} that the CLEO data favor the asymptotic form
rather than the Chernyak-Zhitnitsky one \cite{CZ}, which emphasizes
the endpoint region. The E791 di-jets data support the nearly
asymptotic pion DA above $M_J^2=10$ GeV$^2$ \cite{E791}, $M_J^2$
being the mass squared of the di-jets. A review on the determination
of the leading-twist pion DA can be found in \cite{BMS2}. It seems
that the measurements of the pion transition form factor at low and
high $Q^2$ imply different shapes of the pion DA, which are
contradictory to each other.

In this letter we shall study the process
$\pi\gamma^*\to\gamma$ using the $k_T$
factorization theorem
\cite{CCH,Collins:1991ty,Levin:1991ry,Botts:1989kf,Li:1992nu,HSK}.
If a flat pion DA is favored, it will enhance the
contribution from the region with small momentum fraction $x$.
Once the small $x$ region is important, the parton transverse
momentum squared $k_T^2$, being of the same order as $xQ^2$ in the
internal quark propagator, is not negligible. The regulator $m^2$ in
Eq.~(\ref{Pol}) can be interpreted as $k_T^2$ at LO, whose inclusion
smears the endpoint singularity from $x\to 0$. Besides,
the loop correction to the virtual photon vertex generates the large
double logarithms $\alpha_s\ln^2(Q^2/k_T^2)$ and $\alpha_s\ln^2 x$ in the small $x$
region \cite{Nandi:2007qx}. The former is absorbed into the
$k_T$ dependent wave function, and organized to all orders by
the $k_T$ resummation \cite{Botts:1989kf,CS,Musatov:1997pu}. The
latter is absorbed into a jet function, and organized to all orders
by the threshold resummation \cite{THRE,TMD,FMW1}. With the above
resummation factors, a flat pion DA does not develop an endpoint
singularity.

The NLO correction to the hard kernel of the pion transition form
factor has been computed in the $k_T$ factorization
\cite{Nandi:2007qx}. Fourier transforming Eq.~(44) in
\cite{Nandi:2007qx} into the impact parameter $b_T$ space, we derive
\begin{eqnarray}
F_{\pi\gamma}(Q^2) &=& \frac{\sqrt{2} f_\pi}{3} \int_0^1 dx
\int_0^\infty b_T db_T\, \phi_\pi(x) \exp[-S(x,b_T,Q,\mu)]\, S_t(x, Q)
K_0\left( \sqrt{x}Qb_T \right)
\nonumber\\
&&\hspace{7mm}\times \left[ 1 - \frac{\alpha_s(\mu)}{4\pi}C_F \left(
\ln\frac{\mu^2b_T}{2\sqrt{x}Q} + \gamma_E + 2\ln x
+3-\frac{\pi^2}{3} \right) \right], 
\end{eqnarray}
where $C_F=4/3$ is a color factor and $\gamma_E$ the Euler constant. 
The NLO term in the square brackets implies that the renormalization
scale $\mu$ should be chosen as $\mu^2\sim O(\sqrt{x}Q/b_T)$ in
order to minimize the logarithm. However, we have always set
$\mu=\max(\sqrt{x}Q,1/b_T)$ in our previous analysis of exclusive
processes \cite{Li:1992nu}, so we shall continue to adopt this
choice here. It has been confirmed that the above two choices of
$\mu$ produce almost identical results. The Sudakov exponent from
the $k_T$ resummation is given by
\begin{eqnarray}
S(x,b_T,Q,\mu)=s\left(x\frac{Q}{\sqrt{2}},b_T\right)
+s\left((1-x)\frac{Q}{\sqrt{2}},b_T\right) + 2\int_{1/b_T}^{\mu}
        \frac{d\bar\mu}{\bar\mu}\gamma_q(\alpha_s(\bar\mu)),\label{sud}
\end{eqnarray}
where the explicit expression of the function $s$ and the quark
anomalous dimension $\gamma_q$ can be found in \cite{LY1,KLS}.

\begin{table}[ht]
\begin{center}
\begin{tabular}{cccccc}\hline
$Q$ (GeV) & 2  & 3
 & 4  & 5  & 6  \\\hline
$c$ & 1.7 & 0.75 & 0.47 & 0.32 & 0.25 \\\hline
\end{tabular}
\caption{The power $c$ in the threshold resummation at different
$Q^2$. }\label{tab1}
\end{center}
\end{table}

The power $c\approx 0.3$ in the parametrization $S_t(x, Q)$ for the
threshold resummation \cite{TLS}
\begin{eqnarray}
S_t(x, Q)=\frac{2^{1+2c}\,\Gamma(3/2+c)}
{\sqrt{\pi}\,\Gamma(1+c)}[x(1-x)]^c,\label{thr}
\end{eqnarray}
was derived at the scale of the $B$ meson mass $m_B=5.28$ GeV. The
power-law behavior should be modified as $Q^2$ runs within a large
range in the present case. We determine the $Q^2$ dependence of the
parameter $c$ by repeating the procedure in Appendix D of
\cite{TLS}, which involves the best fit of Eq.~(\ref{thr}) to the
exact resummation formula in the Mellin space $N$ 
\begin{eqnarray}
S_t(N, Q)=\exp\left[\frac{1}{2}\int_0^{1-1/N}\frac{dz}{1-z}
\int_{(1-z)}^{(1-z)^2}
\frac{d\lambda}{\lambda}
\gamma_K\left(\alpha_s\left(\sqrt{\lambda Q^2/2}\right)\right)
\right],
\end{eqnarray}
with the anomalous dimension 
\begin{equation}
\gamma_K=\frac{\alpha_s}{\pi}C_F+\left(\frac{\alpha_s}{\pi}
\right)^2C_F\left[C_A\left(\frac{67}{36}
-\frac{\pi^{2}}{12}\right)-\frac{5}{18}n_{f}\right],
\end{equation}
$n_{f}$ being the number of quark flavors and $C_A=3$ a color factor.
The outcome is
listed in Table~\ref{tab1}, where the rapid increase of $c$ with
decreasing $Q$ arises from the exponentiated radiative correction
proportional to $\alpha_s(Q^2)$. It becomes more difficult to
determine $c$ for $Q < 3$ GeV, because the allowed interval of $N$
shrinks. Therefore, we freeze $c$ at $c=1$, when it exceeds unity.
To simplify the analysis, we propose a parabolic parametrization
\begin{eqnarray}
c\to c(Q^2)=0.04Q^2-0.51Q+1.87\;,\label{power}
\end{eqnarray}
if $c\le 1$. A large value $c\sim 1$, resulting in a quick falloff as
$x\to 0$, will improve the agreement of our prediction with the
low $Q^2$ data, when the flat model is adopted. That is, a careful
treatment of the threshold resummation effect with a running
$c(Q^2)$ is crucial for accommodating the BaBar data in both the low
and high $Q^2$ regions.

We adopt Eq.~(\ref{flat}) for the pion wave function, which is
written as
\begin{eqnarray}
\phi_\pi(x,b_T)=\phi_\pi(x)\exp\left[-\frac{1}{2}\sigma
x(1-x)b_T^2\right]\label{flab}
\end{eqnarray}
in the impact parameter space. The Gaussian form in $b_T^2$ is the
same as proposed in \cite{JKR}, and consistent with that from
\cite{ADSQCD}. For the asymptotic model, $\phi_\pi(x)=6x(1-x)$, the
parameter $\sigma$ can be fixed by the normalization condition
\cite{BHL83}
\begin{eqnarray}
\int_0^1dx\int d^2b_T\, \phi_\pi(x,b_T) = \frac{3}{\pi f_\pi^2}\;,
\end{eqnarray}
giving $\sigma=4\pi^2f_\pi^2=0.677$ GeV$^2$. For the flat model,
$\phi_\pi(x)=1$, the above normalization condition does not apply,
and the parameter $\sigma$ will be tuned to fit the BaBar data. In
the analysis below we choose $\sigma=2.5$ GeV$^2$. Note that the
renormalization-group evolution of the Gegenbauer expansion starting
with the asymptotic form has been known. However, the evolution
starting with the flat form has not yet been studied. Hence, the
models adopted here should be regarded as being defined at a low
normalization point $\mu_0$.

The LO and NLO predictions for the pion transition form factor from
the $k_T$ factorization are presented in Fig.~\ref{fig1}. It is
observed that the curves for $Q^2F_{\pi\gamma}(Q^2)$ from the
asymptotic pion DA saturates quickly for $Q^2>5$ GeV$^2$.
That is, it is difficult to explain the continuous growth of
$Q^2F_{\pi\gamma}(Q^2)$ above $Q^2 \approx 10$ GeV$^2$ using the
asymptotic model. The overshoot in the low $Q^2$ region is expected,
since this portion of data favors an endpoint suppressed shape
\cite{BMS,Guo:2008zzg}. Note that the overshoot can be resolved
within the theoretical uncertainty from varying parameters in our
formalism. For example, multiplying the arguments $xQ/\sqrt{2}$ and
$(1-x)Q/\sqrt{2}$ by a factor 2 in Eq.~(\ref{sud})
\cite{Botts:1989kf} leads to stronger suppression at $x\to 0$, and
better agreement with the data at low $Q^2$. The curves from the
flat pion DA show a good fit to the BaBar data, whose
logarithmic increase with $Q^2$ is a combined consequence of the
inclusion of the parton $k_T$ and the employment of the flat model.
We stress that the fit would deteriorate in the low $Q^2$ region
without a careful treatment of the threshold resummation effect.
To highlight this effect, we compare the
results from the flat model with a constant power $c=0.3$ and with a
running $c(Q^2)$ in Fig.~\ref{fig2}. It is found that the former
exceeds the data around $Q^2\approx 10$ GeV$^2$. The results from
the asymptotic model is almost independent of the running of
$c(Q^2)$, because the asymptotic model decreases quickly enough at
small $x$ by itself. The NLO correction, amounting up to 20\% for
the flat pion DA as shown in Fig.~\ref{fig1}, is under control in
the $k_T$ factorization. This observation is opposite to that
postulated in \cite{Polyakov:2009je}, and to that obtained in the
collinear factorization \cite{Radyushkin:2009zg}.

\begin{figure}[t]
\begin{center}
\includegraphics[height=8.5cm]{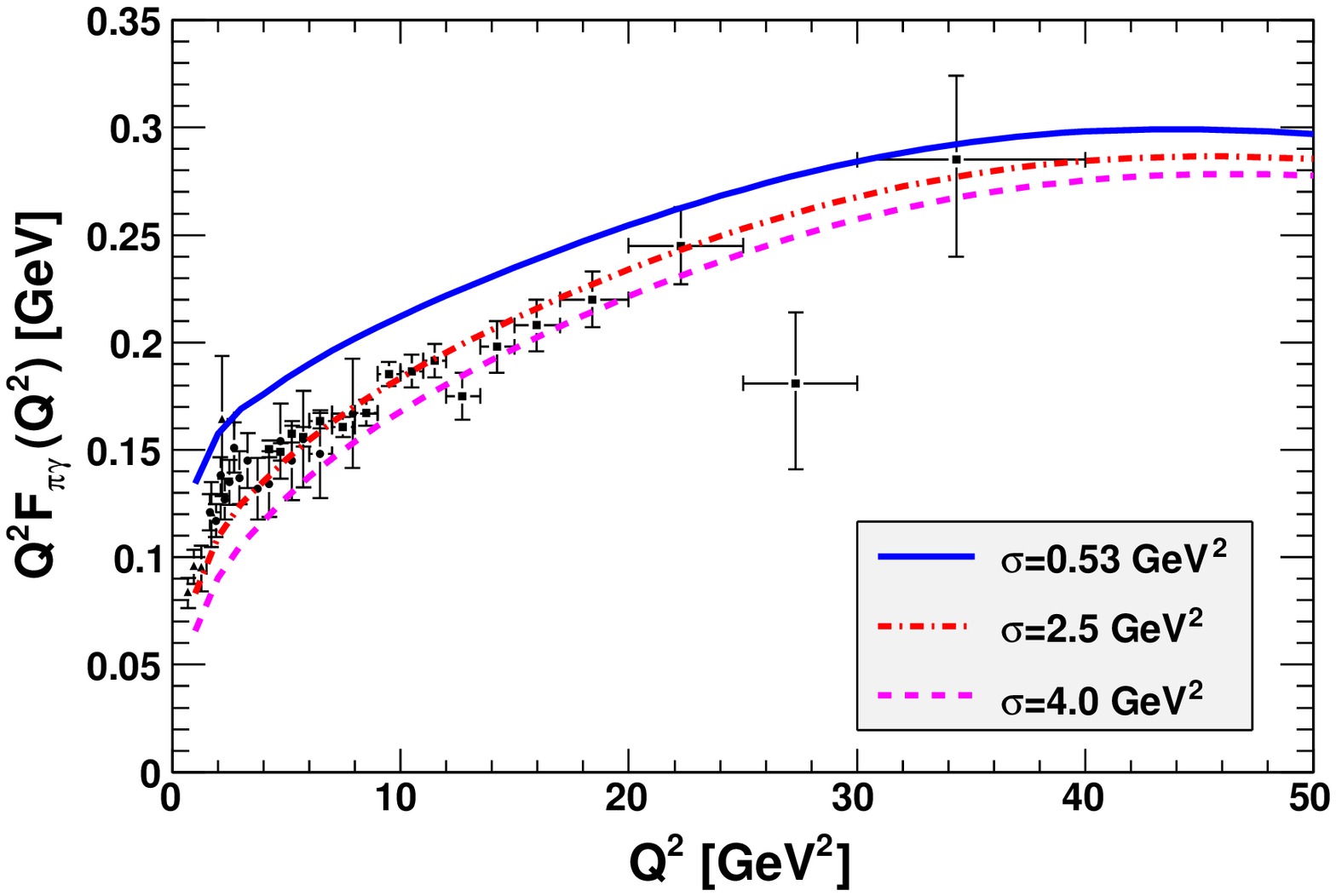}\\
\vspace{-4.5mm}
(a)\\
\includegraphics[height=8.5cm]{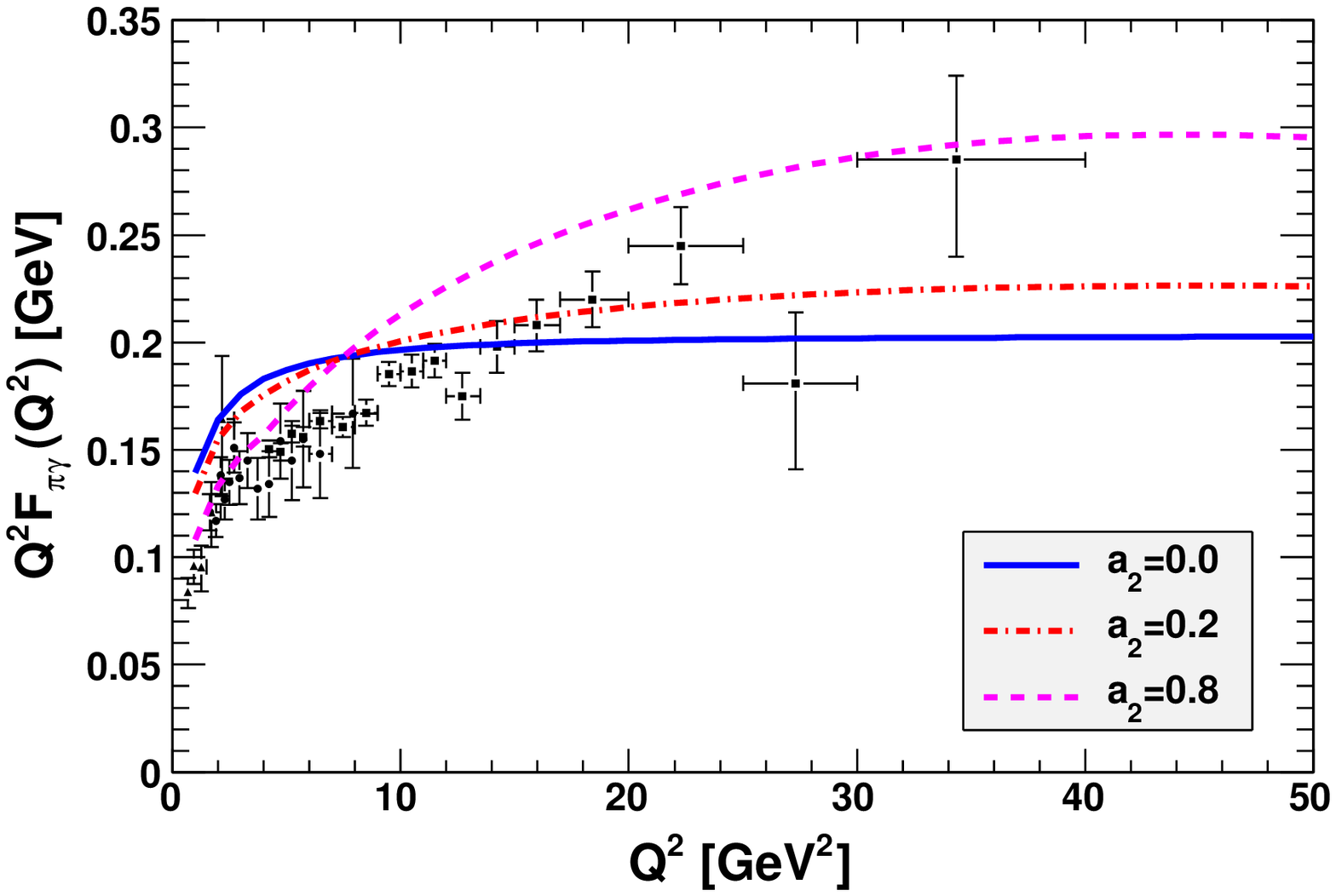}\\
\vspace{-4.5mm} (b) \caption{$Q^2$ dependence of
$Q^2F_{\pi\gamma}(Q^2)$ from (a) the flat model with different $\sigma$.
and (b) Eq.~(\ref{a2}) with different $a_2$.}\label{fig3}
\end{center}
\end{figure}

We should keep in mind that the BaBar data suffer from a large
uncertainty in the high $Q^2$ region. In particular, the data point
for $Q^2=27.31$ GeV$^2$ drops, and is consistent with the asymptotic
limit in Eq.~(\ref{as}). Hence, we examine the dependence of our
prediction on the parameter $\sigma$ associated with the intrinsic
$b_T$ dependence in Eq.~(\ref{flab}). The results are displayed in 
Fig.~\ref{fig3}(a), indicating that the slope of the curves is
insensitive to the variation of $\sigma$ especially in the high
$Q^2$ region. If including the higher Gegenbauer term into the
asymptotic model
\begin{eqnarray}
\phi_\pi(x)= 6x(1-x)\left[ 1+a_2\,C_2^{3/2}(2x-1) \right],\label{a2}
\label{pionDA}
\end{eqnarray}
with the Gegenbauer polynomial $C_2^{3/2}(t)=(3/2)(5t^2-1)$, the
slope of the curves changes significantly with $a_2$ as shown in
Fig.~\ref{fig3}(b). The prediction is enhanced by 10\% for $a_2=0.2$
in the high $Q^2$ region, and saturates around $Q^2\approx 10$
GeV$^2$. Only with an extremely large $a_2\approx 0.8$, which also
emphasizes the endpoint regions, the prediction becomes close to
that from the flat pion DA. It has been mentioned
\cite{Polyakov:2009je} that a very broad pion DA, vanishing at the
endpoints, but with a rapid increase ($\phi_\pi^\prime(0)/6\gg 1$),
serves the purpose of accommodating the BaBar data. However, the big
value $a_2\approx 0.8$ may render the Gegenbauer expansion starting
with the asymptotic model questionable, as claimed in
\cite{Polyakov:2009je}. In summary, the precise measurement of the
pion transition form factor in the high $Q^2$ region will settle
down the issue on the shape of the leading-twist pion DA: if the
growth with $Q^2$ observed currently persists, the flat model is
favored. If the growth becomes milder in the future, the asymptotic model with
nonvanishing higher Gegenbauer terms will be still allowed.

Finally, we comment on the $k_T$ factorization formalism developed
in \cite{TMD}, which involves on-shell partons in the calculation of
hard kernels. The usual procedure of factorizing a $k_T$ dependent
wave function is to neglect the minus component $k^-$ in a hard
kernel, and then to integrate out $k^-$ in the general wave function
that depends on the four components $k^+=xp^+$, $k^-$, and $k_T$,
\begin{eqnarray}
\psi(x,k_T)\equiv\int dk^-\Psi(x,k^-,k_T).
\end{eqnarray}
This is the reason why a parton, participating in the hard scattering,
carries an off-shell momentum $(k^+,0, k_T)$ in our formalism
\cite{Li:2008hu}. The prescription for calculating a $k_T$ dependent
hard kernel, the definition for a $k_T$ dependent wave function, and
the gauge invariance of the $k_T$ factorization have been carefully
addressed in \cite{Nandi:2007qx,Li:2008hu}, which confront the
criticism from \cite{Feng:2008zs,Feng:2008nn}\footnote{The authors
of \cite{Feng:2008nn} rebutted our comments addressed in
\cite{Li:2008hu} on \cite{Feng:2008zs}. They claimed that the $k_T$
factorization theorem for exclusive processes violates gauge
invariance for the reasons: 1) Our method for the one-loop
calculation of Fig.~2(b) in \cite{Li:2008hu} leads to a wave
function nonvanishing for $x>1$, and violates the translation
invariance, since the contribution from $q^+<0$ is non-zero, $q$
being the loop momentum. 2) Our method depends on the contour chosen
in the $q^-$ integration. We disagree with both of them. In our
method, the $q^+<0$ contribution comes only from the limit $q^+\to
0$ from the left, which corresponds to $x\to 1$, and does not
violate the translational invariance.
As for the latter, the authors of \cite{Feng:2008nn} missed the
point of including the semicircles, whose purpose is to avoid the
ambiguity from $q^+\to 0$, when the double pole
$q^-=(q_\perp^2-i\varepsilon)/(2q^+)$ moves to the infinity. If
choosing the lower semicircle, the poles will cross the semicircle
as $q^+\to 0$, and the ambiguity comes back. Besides, the authors of
\cite{Feng:2008nn} relied on the claim that all contributions from
the semicircles in the $q^-$ contour integration vanish in the limit
of the semicircle radius $R\to \infty$, but it is incorrect. As
explained in \cite{Li:2008hu}, these contributions do not vanish as
$q^+\to 0$. Note that the same problem has been discussed in
literature, {\it e.g.} in \cite{TMYan}. Moreover, the reply in
\cite{Feng:2008nn} is selective; {\it e.g.} there was no answer to
our question raised in \cite{Li:2008hu}: {\it why is the singularity
from a highly off-shell gluon with $q^2\to\infty$ a light-cone
singularity?}}. The transverse momentum squared $k_T^2$ appears in
the hard kernel for the pion transition form factor through the
internal quark invariant mass, $(k-P_\gamma)^2=-xQ^2-k_T^2$, for the
final-state photon momentum $P_\gamma=(0,P_\gamma^-,0_T)$ and
$k^2=-k_T^2$. Assuming on-shell partons, the hard kernel is
independent of $k_T^2$ because of $(k-P_\gamma)^2=-xQ^2$ for
$k^2=0$. Therefore, it will be difficult to explain the BaBar data
using the formalism of \cite{TMD}.

We thank Ahmed Ali for useful comments. 
This work was supported by the National Center for Theoretical
Sciences and National Science Council of R.O.C. under Grant No.
NSC-95-2112-M-050-MY3.

\end{document}